# Unusual negative formation enthalpies and atomic ordering in isovalent alloys of transition metal dichalcogenide monolayers

Ji-Hui Yang and Boris I. Yakobson*

Department of Materials Science and Nanoengineering, Rice University, Houston, Texas 77005, USA

**Abstract:** Common substitutional isovalent semiconductor alloys usually form disordered metastable phases with positive excess formation enthalpies (ΔH). In contrast, monolayer alloys of transition metal dichalcogenides (TMDs) $MX_2$ (M = Mo, W; X = S, Se) always have negative ΔH, suggesting atomic ordering, which is, however, not yet experimentally observed. Using first-principles calculations, we find that the negative ΔH of cation-mixed TMD alloys results from the charge transfer from weak Mo-X to nearest strong W-X bonds and the negative ΔH of anion-mixed TMD alloys comes from the larger energy gain due to the charge transfer from Se to nearest S atoms than the energy cost due to the lattice mismatch. Consequently, cation-mixed and anion-mixed alloys should energetically prefer to have Mo-X-W and S-M-Se ordering, respectively. The atomic ordering, however, is only locally ordered but disordered in the long range due to the symmetry of TMD monolayers, as demonstrated by many energetically degenerate structures for given alloy compositions. Besides, the local ordering and disordering effects on the macroscopic properties such as bandgaps and optical absorptions are negligible, making the experimental observation of locally ordered TMD alloys challenging. We propose to take the advantage of microscopic properties such as defects which strongly depend on local atomic configurations for experiments to identify the disordering and local ordering in TMD alloys. Finally, quaternary TMD alloys by mixing both cations and anions are studied to have a wide range of bandgaps for optoelectronic applications. Our work is expected to help the formation and utilization of TMD alloys.

Substitutional isovalent semiconductor alloy such as $A_{1-x}B_xC$ has greatly advanced semiconductor science and technologies during the past decades, as they can achieve continuously tunable material properties between the pure end-point constituents AC and BC by varying the composition x, thus broadening the range of available material properties for extensive applications, such as Si1-x$Ge_x$ for thermoelectrics,[1,2] $Al_{1-x}Ga_xAs$ for quantum wells,[3] $CuIn_{1-x}Ga_xSe_2$ for solar cells, $In_{1-x}Ga_xN$ for light emitting diodes,[5] and $Hg_{1-x}Cd_xTe$ for infrared detectors.[6] Usually, isovalent semiconductor alloys form disordered (D) metastable phases with positive excess enthalpies of formation $\Delta H_D(x) = E(A_{1-x}B_xC) - (1-x)E(AC) - xE(BC)$, where E(AC), E(BC), and $E(A_{1-x}B_xC)$ are the total energies of pure AC, pure BC, and their mixed alloys, respectively.[7] This positiveness of $\Delta H$ mainly results from the strain energy attendant upon mixing two end-point constituents with dissimilar lattice constants, and thus random isovalent semiconductor alloys usually form at above a critical temperature when the entropic energy gain is larger than $\Delta H$.[7] Besides, isovalent semiconductor alloys can also form some special ordered (O) structures with unique topological property to possess just enough structural degrees of freedom to accommodate any bond length and bond angle, thus lowering energies by minimization of strain.[8,9] However, the formation enthalpies of ordered alloys $\Delta H_O(x)$ are still often positive[7,10] except very few examples in the tetragonal alloys such as SiC in the zinc-blende structure[8] and $AlInX_2$ (X = P, As) in the chalcopyrite structure.[11] As a result, the ground state of periodic, isovalent bulk semiconductor alloys $A_{1-x}B_xC$ is generally phase separation into (1-x)AC +(x)BC.[7]

Things become different in the cases of monolayer isovalent alloys of transition metal dichalcogenides (TMDs) $MX_2$ (M = Mo, W; X = S, Se) —— they always have negative $\Delta H$. The recent theoretical study of using the ideal random isovalent alloys of TMD monolayers such as $Mo_{1-x}W_xSe_2$ to suppress deep defect levels shows that TMD isovalent alloys have negative $\Delta H$ although the isovalent cations are totally disordered.[12] The negativeness of $\Delta H$, however, suggests that either short or long-range atomic ordering should exist in $Mo_{1-x}W_xSe_2$ according to Hume-Rothery rules.[7] Indeed, Gan *et al.* theoretically reported ordered cation-mixed TMD monolayer isovalent alloys at x = 1/3 and 2/3 with energetic preference of Mo-X-W over Mo-X-Mo and W-X-W interactions.[13] Kutana *et al.* also predicted many thermodynamic ordered ground states of $Mo_{1-x}W_xS_2$ alloys with negative ΔH.[14] Similar results are also reported in anion-mixed TMD monolayer isovalent alloys. For example, Komas *et al.* reported negative ΔH of $Mo(S_{1-x}Se_x)_2$ alloys in which short-range atomic order favors having different anion atoms in the nearest neighbor sites in the chalcogen sublattice.[15] Kang *et al.* further reported negative ΔH of ideal random anion-mixed alloys and ordered alloy structures of $Mo(S_{1-x}Se_x)_2$ at x = 1/3, 1/2, and 2/3.[16] The unusually negative ΔH of both ordered and disordered monolayer isovalent alloys of TMDs, however, still lacks clear explanations despite it's believed to be related to the joint effect of volume deformation, chemical difference, and a low-dimensionality enhanced structure relaxation.[16] Besides, it's also not clear how the atomic ordering or disordering will affect the properties of TMD monolayer isovalent alloys, especially those that are sensitive to local atomic configurations, such as defect properties. As TMD alloys are now attracting more and more interests with many experimental works reported,[17-32] understanding the above questions will be important for the formation and further utilization of TMD alloys.

In this work, using first-principles calculation methods, we study the formation and atomic ordering of TMD monolayer isovalent alloys. The unusual negative ΔH of TMD alloys is explained as follows: the unusually negative ΔH of cation-mixed TMD alloys, either disordered or ordered, results from the charge transfer from weak Mo-X bonds to nearest strong W-X bonds due to the negligible lattice mismatch between common-anion TMDs; for anion-mixed alloys, ΔH can be positive or negative due to the relative large lattice mismatch between common-cation TMDs and the negative ΔH comes from the larger energy gain due to the charge transfer from Se atoms to nearest S atoms than the energy cost due to the lattice mismatch. Consequently, energetically cation-mixed alloys should prefer to have Mo-X-W ordering and anion-mixed alloys should prefer to have S-M-Se ordering. However, this kind of local ordering is yet not experimentally confirmed. Besides the common explanation that the disordered TMD alloys are more energetically stable due to the largest entropic energy gain at the growth temperatures, we propose two other possible reasons. First, we find that atomic ordering in TMD alloys is only limited within short range but doesn't have to be long range due to the symmetry of TMD monolayers. Therefore, there can be many energetically degenerate structures for a given composition x, which might make the experimental observation of ordered TMD alloys difficult. Second, we study the local ordering and disordering effects on the properties of TMD alloys. Generally, their effects on the macroscopic properties such as band gaps and optical absorptions are negligible, further making the experimental observation of ordered TMD alloys challenging. However, for microscopic properties such as defects which strongly depend on local atomic configurations, the local ordering will suppress some atomic configurations and thus can strongly affect the formation of some defects. As a result, we propose to take the advantage of defect properties for experiments to identify the disordering and local ordering in TMD alloys. Finally, we show the electronic properties of TMDs can be further expanded by forming quaternary alloys with both cations and anions isovalently mixed.

## RESULTS AND DISCUSSION

To find possible ordered ground structures of TMD alloys indicated by negative $\Delta H$, we use cluster expansion (CE) methods[33] in conjunction with first-principles calculations, which has demonstrated high efficiency in dealing with alloy problems.[14,16,34-36] To study the properties of disordered random TMD alloys, we adopt special quasirandom structure (SQS) methods.[37,38] Detailed calculation methods can be found in the Experimental Section. The CE simulation results of four TMD ternary alloys are shown in Figs. 1(a)-(d). Thermodynamically stable ground structures with negative formation enthalpies of several meV are found for both cation-mixed and anion-mixed alloys, in agreement with previous works.[12-16] These ground structures are shown in the Supporting Information. The fitted effective cluster interaction (ECI) parameters, as listed in Table I, show that only some small-size clusters (Fig. 1e) have significant contributions to the formation enthalpies. For the cation-mixed alloys, the nearest pair cluster (2, 1) in the triangle sublattice has the dominant ECI parameter over all the other clusters. The positiveness of ECI parameter for this cluster indicates that Mo likes to stay with W in the neighbor in the cation triangle sublattice, which can give negative nearest pair cluster correlation function (see calculation methods). Similarly, for the anion-mixed alloys, the nearest pair clusters (2, 1) and (2, 2) dominant over the other clusters with positive ECI parameters, indicating that S likes to stay with Se in the neighbor in the anion sublattice. Consequently, the more negative for the correlation functions of the nearest pair clusters, the lower the alloy formation enthalpies are. Note that, for disordered alloys, the pair correlation function, which is $(2x-1)^2$, can never be negative.[37,38] As a result, for the TMD alloys of $MX_2$ (M = Mo, W; X = S, Se), the disordered alloys will always have larger formation enthalpies than those alloys with negative pair correlation functions. To confirm this point, we construct SQS structures that have perfect-matched nearest pair correlation functions compared to random alloys, as shown in Table II. Our total energy calculations demonstrate that the disordered alloys indeed have larger formation enthalpies than the alloys with local ordering of mixed atoms, which have smaller nearest pair correlation functions (see Table II).

However, the negative formation enthalpies of TMD alloys are still not explained so far. To understand this, we consider three-step process during the formation of an isovalent alloy:[39,40] (i) for the volume deformation (VD) contributions, we compress or expand the end-point constituents from their equilibrium lattice constants to the alloy lattice constants; (ii) for the charge-exchange (CEX) term, we mix the two end-point constituents on perfect lattice sites at the alloy lattice constants; (iii) to gain the magnitude of the structural relaxation (SR), we relax all the atomic positions inside the cell using the quantum-mechanical forces. The total formation enthalpies can therefore be decomposed into these three contributions, i.e., $\Delta \mathrm{H} = \Delta \mathrm{H}_{\mathrm{VD}} + \Delta H_{CEX} + \Delta H_{SR}$. Table III lists the decomposition results for $Mo_{1-x}W_xS_2$ and $Mo(S_{1-x}Se_x)_2$ alloys with local atomic ordering. For cation-mixed alloys, the contributions of VD and SR terms to the formation enthalpies are negligible due to the nearly perfect matched lattice constants of $MoS_2$ and $WS_2$. Instead, the contribution of CEX term is dominant and makes ΔH negative. For anion-mixed cases, although the contributions of VD and SR terms are larger than that of the CEX term, ΔH can't be negative without the contribution of CEX term. Same conclusions hold for disordered alloys.

To understand why CEX term is negative, we consider the charge transfer during the mixing process. Fig. 2(a) displays the corresponding difference in charge densities $\Delta\rho^{\mathrm{CEX}}(r) = \rho(\mathrm{Mo_{0.75}W_{0.25}Se_2}) - 0.75\rho(\mathrm{MoSe_2}) - 0.25\rho(\mathrm{WSe_2})$, which gives the charge transfer information during the formation of $\mathrm{Mo_{0.75}W_{0.25}Se_2}$ alloys. Here, $\mathrm{Mo_{0.75}W_{0.25}Se_2}$, $\mathrm{MoSe_2}$, and $\mathrm{WSe_2}$ are in the same ideal lattice of the alloy. As clearly seen, electrons are transferred from Mo-Se bonds to W-Se bonds after mixing $MoSe_2$ and $WSe_2$. Because $WSe_2$ has a larger cohesive energy than $MoSe_2$, i.e., 15.57 eV per formula for $WSe_2$ versus 13.82 eV per formula for $MoSe_2$, this process thus has energy gain, resulting in the negative $\Delta H_{CEX}$. Because the lattice mismatch is negligible for common-anion TMDs and energy is always gained during the CEX process of alloy formation, ΔH is always negative no matter how the cations are arranged, as shown in Figs 1(a) and 1(b). Note that, different cation arrangements can still affect the charge transfer strength and thus the total energy gain, resulting in the variation of ΔH.

For the anion-mixed cases, the charge density difference during CEX process, i.e., $\Delta\rho^{\mathrm{CEX}}(r) = \rho[\mathrm{Mo(S_{0.66}Se_{0.33})_2}] - 0.66\rho(\mathrm{MoS_2}) - 0.33\rho(\mathrm{MoSe_2})$, gives the charge transfer information during the formation of $Mo(S_{0.66}Se_{0.33})_2$. Fig. 2(b) shows that electrons are transferred from Se to S after mixing $MoS_2$ and $MoSe_2$, although there is also some electron accumulation at the Mo-Se bonds. Again, because $MoS_2$ has a larger cohesive energy than $MoSe_2$ (15.43 eV per formula for $MoS_2$ versus 13.82 eV for $MoSe_2$), extra energy is gained in this process, resulting in the negative $\Delta H_{CEX}$. However, the common-cation TMDs generally have considerable lattice mismatch and when the anions are mixed, the net contributions of VD and SR terms to ΔH are significant and positive. Depending on the anion arrangements, the negativness of $\Delta H_{CEX}$ can be large or small and thus ΔH can be positive or negative, as seen in Figs 1(c) and 1(d). Only with local ordering and thus enough S-M-Se connections, can $\Delta H_{CEX}$ be negative enough to make ΔH negative.

The negative formation enthalpies of TMD alloys indicate that these alloys, either disordered or locally ordered, can even stably exist at T = 0 K. As locally ordered alloys with larger energy gain in the CEX process are more energetically preferred than disordered ones, they should be more likely to form, especially at low temperatures. However, the locally ordered TMD alloys have not been experimentally confirmed so far. One common explanation is that at the growth temperatures, the disordered TMD alloys with the largest entropic energy gain can be more energetic favorable than those locally ordered alloys. However, the entropic energy difference between disordered and locally ordered TMD alloys is not known and difficult to be determined from both experimental and theoretical aspects. Here, we consider other possible factors that could make the experimental observation of locally ordered TMD alloys more challenging.

One of the important factors is the lack of long range atomic ordering in TMD alloys. Our above analysis shows that, cation-mixed TMD alloys with the maximum number of Mo-X-W connections tend to have the lowest formation enthalpies. In TMDs, each cation has six nearest neighbors in the cation triangle sublattices with $C_{6v}$ symmetry. In the dilute mixed case with a very small or large x, the number of Mo-X-W connections can be maximized to be six times of the number of minor mixed cations by arranging the minor mixed cations all surrounded by the major mixed ions as the nearest neighbors in the cation sublattices. Apparently, there can be many such atomic configurations, which are symmetry-nonequivalent and energetically degenerate. In the dense mixed case, i.e., with x = 0.5, the two kinds of mixed cations will inevitably have the same kind of mixed ions as the nearest neighbors due to the $C_{6v}$ symmetry. For example, one can imagine that if a Mo atom is surrounded by six W atoms, then each of the W atoms will have at least two other W atoms at the nearest neighbor sites. Consequently, the triangle lattice with $C_{6v}$ symmetry can tolerant flexible arrangements of mixed cations to maximize the number of Mo-X-W connections, thus allowing many alloy atomic configurations which are energetically degenerate. Similarly, for anion-mixed alloys, many energetically degenerated atomic structures can co-exist for both dilute and dense mixed cases. We demonstrate the energetic degeneracy in dense mixed TMD alloys such as $Mo_{0.5}W_{0.5}S_2$ and $Mo(S_{0.5}Se_{0.5})_2$ in 36-atom supercells. By considering the nearest pair cluster correlation functions, we find 22 and 24 nonequivalent atomic configurations (Supporting Information) with energy difference less than 0.5 meV per mixed atom for $Mo_{0.5}W_{0.5}S_2$ and $Mo(S_{0.5}Se_{0.5})_2$ alloys, respectively. Note that, the larger the supercell, the more energetically degenerate atomic configurations can be found. The energetic degeneracy suggests that, in reality TMD alloys tend to be disordered in the long range even if keeping the local ordering. The long range disordered behavior can mess up atomic configurations of TMD alloys, making it difficult for the experiments to confirm the local ordering.

Another factor that makes the experimental observation of locally ordered TMD alloys challenging is that, the local ordering almost has negligible effects on the macroscopic properties of TMD alloys compared to the disordering. For example, Fig. 3 shows the bandgaps and band edge positions as functions of composition x in the four kinds of TMD alloys. As seen that, at each composition x, the locally ordered TMD alloys (represented by ordered structured obtained in CE simulations) almost have the same values of bandgaps and band edge positions as the disordered alloys (represented by SQS structures), with the largest difference less than 0.05 eV. We further consider the ordering effect on optical properties. Fig. 4 shows the optical absorbance of the four TMD alloys at x = 0.5. Again, no big difference is induced by local atomic ordering compared to the disordering cases.

To distinguish the local ordering in TMD alloys, attentions must be paid to those microscopic properties which are strongly dependent on local atomic configurations. Defect property is one of such properties, as both the defect formation energies and defect levels are strongly related to the local environment due to defect wave-function localization and the size-mismatch-induced strain effect.[12,41] In disordered and locally ordered TMD alloys, defects can have different local surrounding motifs. For example, in the disordered cation-mixed TMD alloy $Mo_{0.75}W_{0.25}Se_2$ (represented by the SQS structure shown in Fig. 5a), any kind of first-neighbor motifs around the defect Se vacancy, is possible, including three Mo atoms ($V^{SQS}_{Se-3Mo}$), two Mo and one W atoms ($V^{SQS}_{Se-2Mo1W}$), one Mo and two W atoms ($V^{SQS}_{Se-1Mo2W}$), or three W atoms ($V^{SQS}_{Se-3W}$). However, in the local ordering case (represented by the ordered structure shown in Fig. 5b), Se vacancy can only have two kinds of first-neighbor motifs, either being three Mo atoms ($V^{order}_{Se-3Mo}$) or two Mo and 1 W atoms ($V^{order}_{Se-2Mo1W}$). Other motifs are avoided to maximize the Mo-Se-W connections in $Mo_{0.75}W_{0.25}Se_2$ ordered alloys. The consequence is that, in disordered $Mo_{0.75}W_{0.25}Se_2$, four kinds of defect states, including a defect state much closer to the conduction band minimum (CBM) created by $V^{SQS}_{Se-3W}$, can be detected, as seen in Fig. 5(c). Because $V^{SQS}_{Se-3W}$ has the smallest formation energy among the four kinds of Se vacancies,[12] it should have the largest defect concentration, making it more distinguishable. In contrast, in ordered $Mo_{0.75}W_{0.25}Se_2$, only two kinds of Se vacancies can be detected and the one with shallow levels close to the CBM is missing, as shown in Fig. 5c. As anion vacancies are common defects and can relatively easily form, experimental detections of different kinds of anion vacancies in TMD alloys are feasible and called for, which can not only verify whether the local ordering exists in TMD alloys, but also test if the alloy strategy proposed in Ref. 12 works to make defect levels shallow.

As both isovalent-cation and isovalent-anion TMD ternary alloys, either disordered or locally ordered, can be easily formed with negative formation enthalpies, isovalent quaternary TMD alloys by mixing both cations and anions are expected, which will not only expand TMD properties more widely but also make the tuning of TMD properties more flexibly. In fact, recent experiments have realized the synthesis of quaternary TMD alloys.[42] Because the ordering effect on the macroscopic properties is negligible, we use SQS structure models to study the bandgaps of quaternary TMD alloys $Mo_{1-x}W_x(S_{1-y}Se_y)_2$ as function of compositions x and y. Our calculation results in Fig. 6 show that the bandgaps of quaternary TMD alloys can be continuously tuned from about 1.40 eV to more than 1.80 eV, which lie in the range of optimal bandgaps for solar cell absorbers according to Shockley–Queisser theory. Consequently, TMD alloys can be strong candidates for optoelectronic applications.

## CONCLUSION

In conclusion, we have explained the unusual negative formation enthalpies of TMD alloys. We find that the negative ΔH of cation-mixed TMD alloys results from the charge transfer from weak Mo-X bonds to nearest strong W-X bonds and the negative ΔH of anion-mixed TMD alloys comes from the larger energy gain due to the charge transfer from Se atoms to nearest S atoms than the energy cost due to the lattice mismatch. Consequently, cation-mixed alloys prefer to have Mo-X-W ordering and anion-mixed alloys prefer to have S-M-Se ordering. We further find that atomic ordering in TMD alloys is only locally ordered but disordered in the long range due to the symmetry of TMD monolayers, as there are many energetically degenerate structures for a given composition. Besides, we find the local ordering and disordering effects on the macroscopic properties such as band gaps and optical absorptions of TMD alloys are negligible. These things will make the experimental observation of ordered TMD alloys difficult and challenging. We propose to take the advantage of microscopic properties such as defects which strongly depend on local atomic configurations for experiments to identify the disordering and local ordering in TMD alloys. Finally, we show that quaternary TMD alloys with both cations and anions isovalently mixed can further expand the electronic properties of TMDs for optoelectronic applications. Our work is expected to help the formation and utilization of TMD alloys. We note that, a recent experimental work[43] does observe more (S, Se) than (Se, Se) vertical stacking in $W(S_{1-x}Se_x)_2$ monolayer alloys, supporting the local atomic ordering and in agreement with our theoretical results.

## EXPERIMENTAL SECTION

**Alloy simulation methods.** The possible ground states of TMD alloys with local atomic ordering are obtained by using CE formalism,[33] which provides an effective way to sample a $2^N$-dimensional configuration space of a binary alloy in an N-site well-defined lattice. In the CE formalism, any alloy configuration can be represented by a vector $\boldsymbol{\sigma}$ containing a set of spin $\sigma_i$ (i = 1, 2, …, N), with $\sigma_i$ being equal to +1 or -1 for one or the other kind of mixed atoms. Then any physical property of a specific configuration, such as total energies and band gaps, can be expanded as:[44]

$$\mathrm{f}(\boldsymbol{\sigma}) = \sum_{\alpha} m_\alpha J_\alpha \langle \prod_{i\in\alpha'} \sigma_i \rangle \qquad (1)$$

where α is a cluster, $m_\alpha$ is the number of symmetry-equivalent clusters of α, $J_\alpha$ is the ECI parameters for the given physical property and the summation is taken over all symmetry-nonequivalent clusters. Angle bracket designates a lattice average of the spin product of symmetry-equivalent clusters $\alpha'$ of α over all the N sites, which is also called cluster correlation function. Using the information obtained from first-principles calculations of certain number of artificially chosen alloy configurations, the ECI parameters $J_\alpha$ can be fitted through Eq. (1) and then the corresponding properties of many more structures can be fast obtained by just calculating the angle bracket term. Often, cluster α can be denoted as (k, m), e.g., pairs of atoms (a figure with k = 2 vertices separated by an mth neighbor distance), triangles (k = 3 vertices).[37,38] The advantage of CE is that Eq. (1) converges rapidly with (k, m) and it is often sufficient enough to use just several small-size clusters to get very accurate physical property $\mathrm{f}(\boldsymbol{\sigma})$ of configuration $\boldsymbol{\sigma}$. Besides, according to CE formalism, two structures should have the same physical properties if they have the same cluster correlation function. For random isovalent alloys $A_{1-x}B_xC$, the correlation function for (k, m) is known as $(2x-1)^k$.[37,38] If a special structure can be constructed to have the same (or very close) correlation functions for every clusters (k, m), then this special structure can be used to study the properties of totally disordered alloys. The idea is known as special quasi-random structure (SQS) method,[37,38] which has been widely used to study random alloys. In this work, the CE fitting of formation enthalpies were carried out with the Alloy-Theoretic Automated Toolkit (ATAT) code.[44] For each alloy, more than 90 ordered structures are used to fit $J_\alpha$. The total number of atoms reaches up to 42 for cation-mixed alloys and to 27 for anion-mixed alloys. The quality of the CE fit is evaluated using a cross-validation score,[45] which is less than 0.5 meV for every CE simulation. To study the disordering random alloys, SQS structures are constructed in $8 \times 8$ supercells for x=1/4, 1/2, 3/4 and in $9 \times 9$ supercells for x=1/3, 2/3 with cluster correlation functions to well match those of random alloys.

**First-principles calculation methods.** First-principles calculations are performed using density-functional theory (DFT)[46,47] as implemented in the VASP code[48,49]. The electron and core interactions are included using the frozen-core projected augmented wave (PAW) approach.[50] and the generalized gradient approximation (GGA) formulated by Perdew, Burke, and Ernzerhof (PBE)[51] is adopted. The structures are relaxed until the atomic forces are less than 0.001 eV/Å and total energies are converged to $10^{-6}$ eV with the cutoff energy for plane-wave basis functions set to 350 eV.

## AUTHOR INFORMATION

**Corresponding author**

E-mail: biy@rice.edu

**Notes**

The authors declare no competing financial interest.

## ACKNOWLEGEMENTS

This work was funded by XXXXX.

## Figures

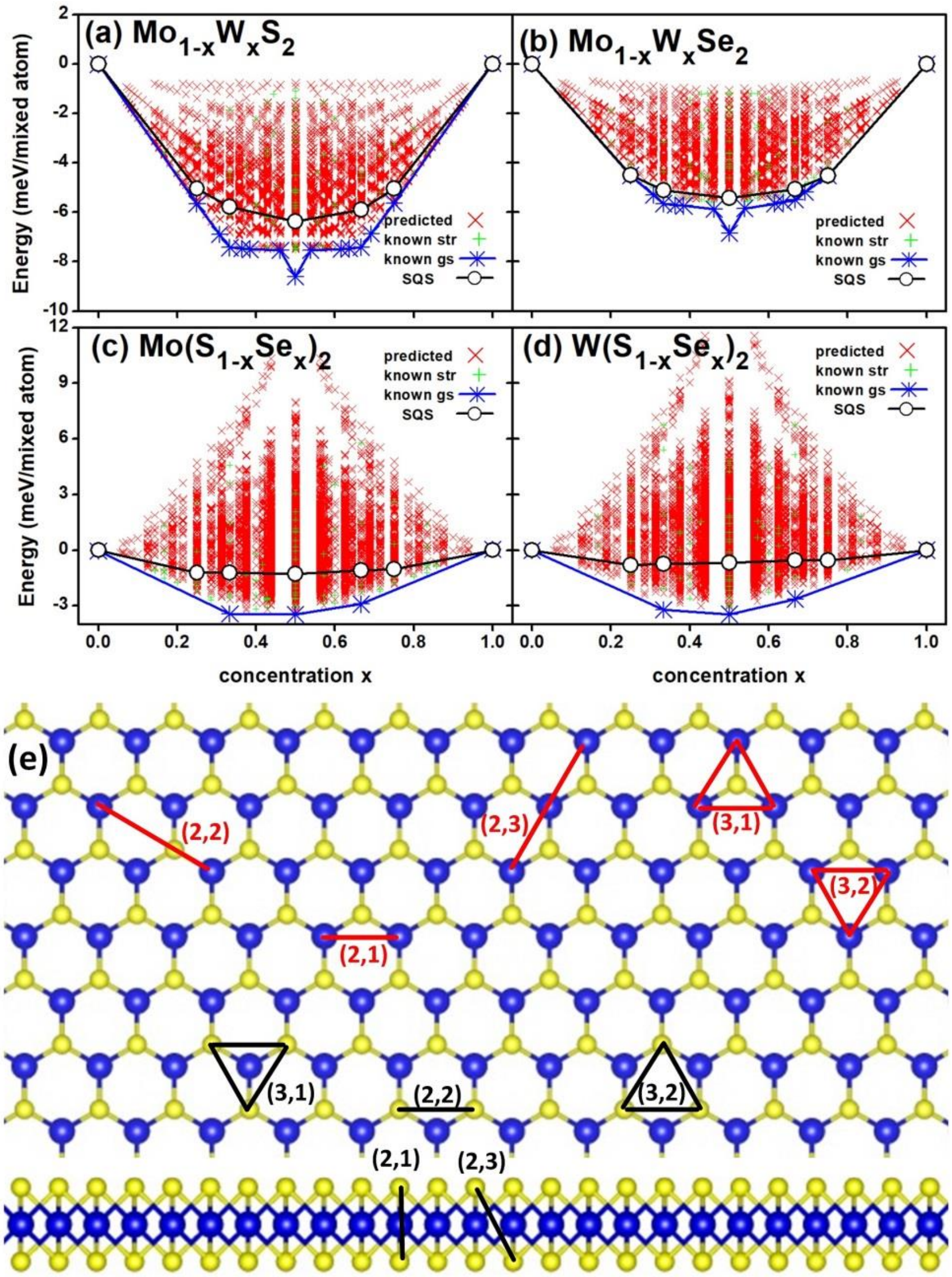

Figure 1. (a)-(d) Cluster expansion simulation results for the formation enthalpies of four TMD ternary alloys at different concentrations. Formation enthalpies of predicted ground structures with local ordering are linked by blue lines. Formation enthalpies of disordered structures (SQS) are also shown (black lines) for comparisons. (e) Dominate cation clusters (red lines) (k, m) and anion clusters (blue lines) in CE simulations. Yellow atoms are S and blue atoms are Mo.

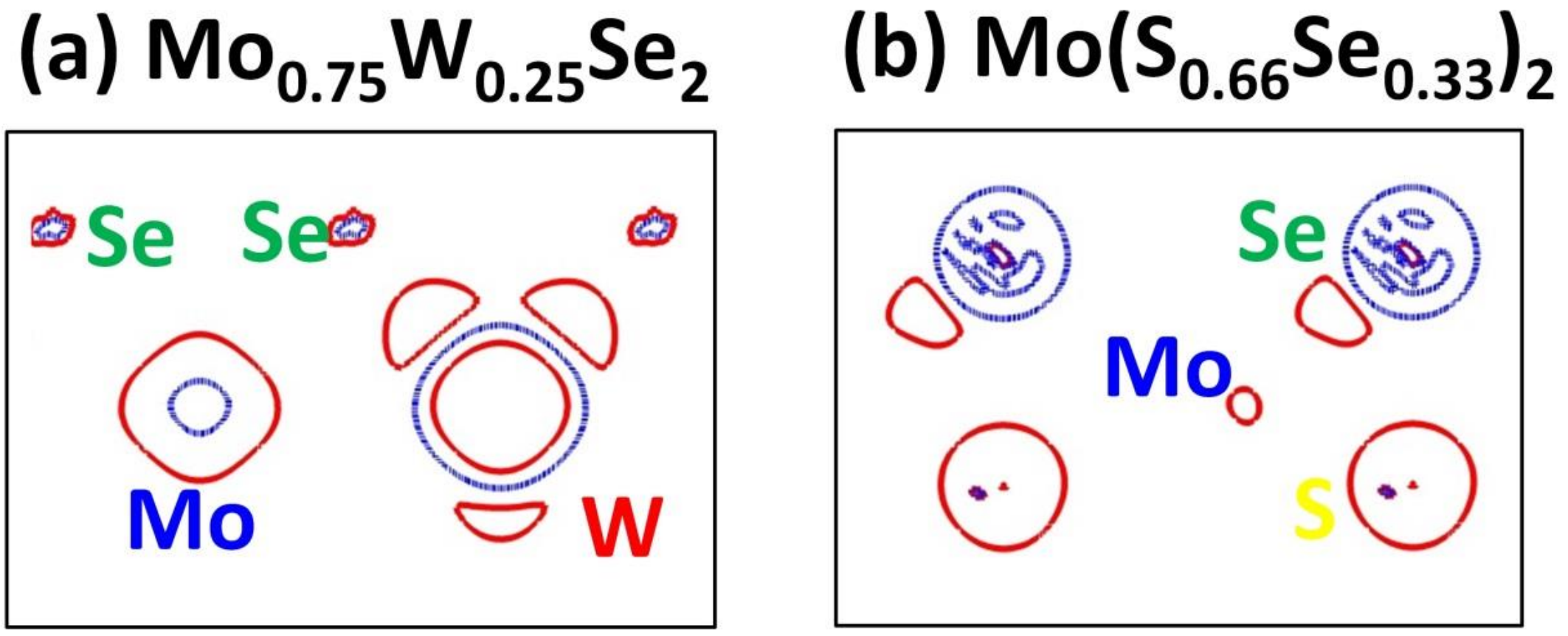

Figure 2. Charge density difference during charge exchange process for (a) $Mo_{0.75}W_{0.25}Se_2$ and (b) $Mo(S_{0.66}Se_{0.33})_2$ alloys. Red solid lines indicate electron accumulations and blue dashed lines indicate electron loss.

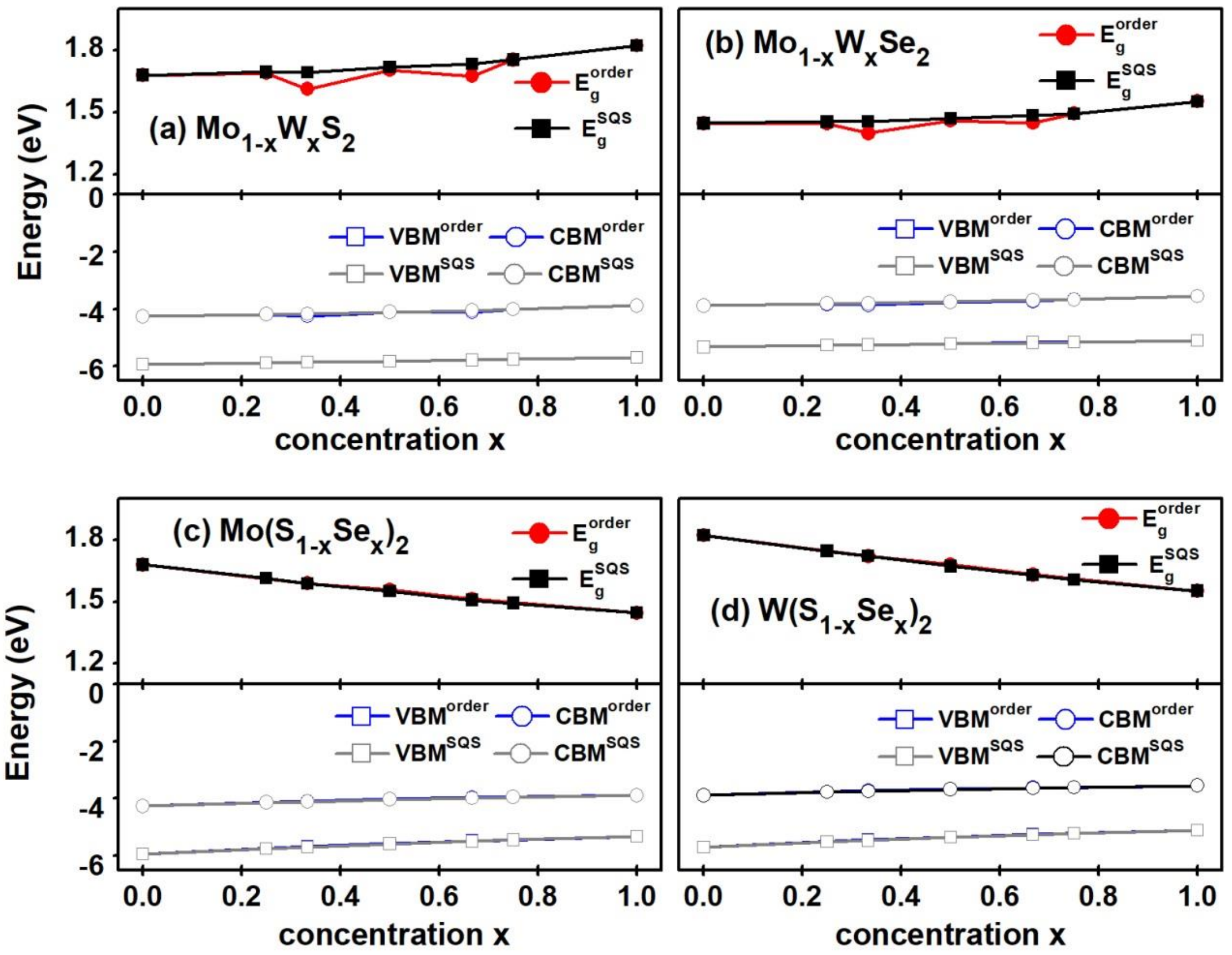

Figure 3. (a)-(d) Bandgaps and band edge positions (referenced to the vacuum level) of four TMD ternary alloys at different concentrations. Values of both the locally ordered alloys (represented by CE predicted ground structures) and random disordered alloys (represented by SQS structures) are given for comparisons.

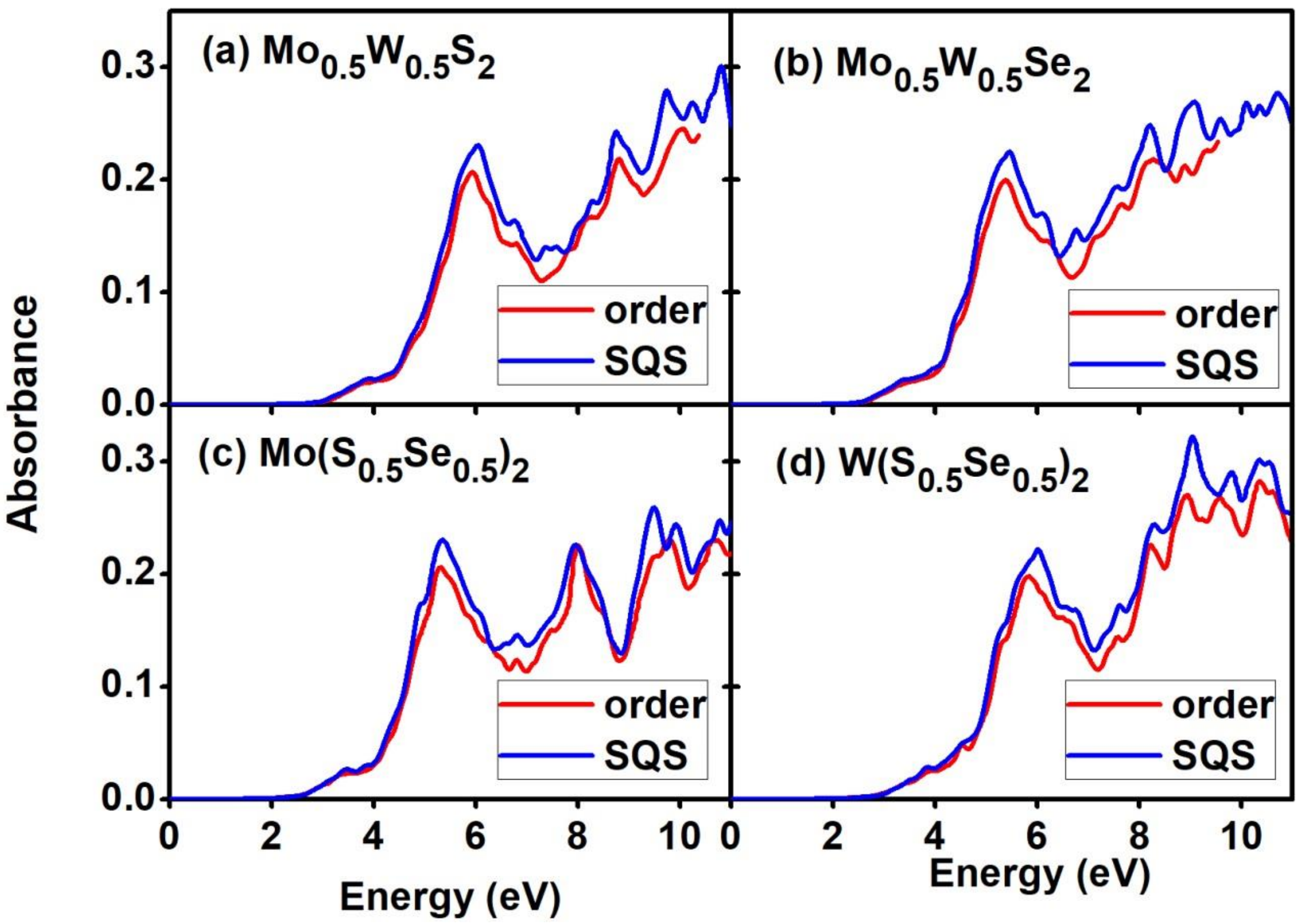

Figure 4. (a)-(d) Optical absorbance of four representative TMD ternary alloys (x = 0.5). The absorbance of both the locally ordered alloys and random disordered alloys are given for comparisons.

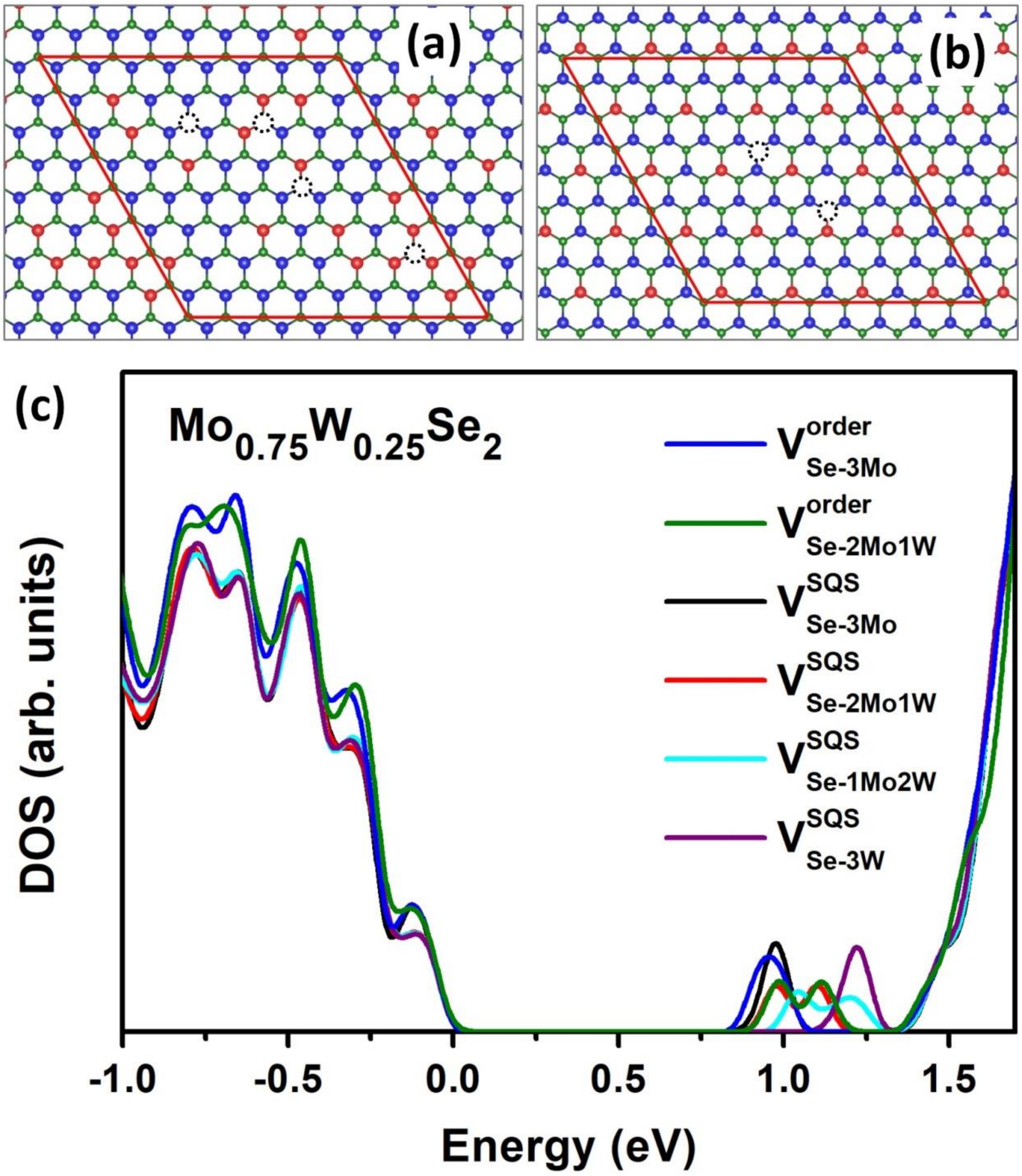

Figure 5. Possible atomic configurations around Se vacancies in (a) disordered and (b) locally ordered $Mo_{0.75}W_{0.25}Se_2$. (c) Density of states of disordered and locally ordered $Mo_{0.75}W_{0.25}Se_2$ alloys with different kinds of Se vacancies. The dotted circles denote Se vacancies.

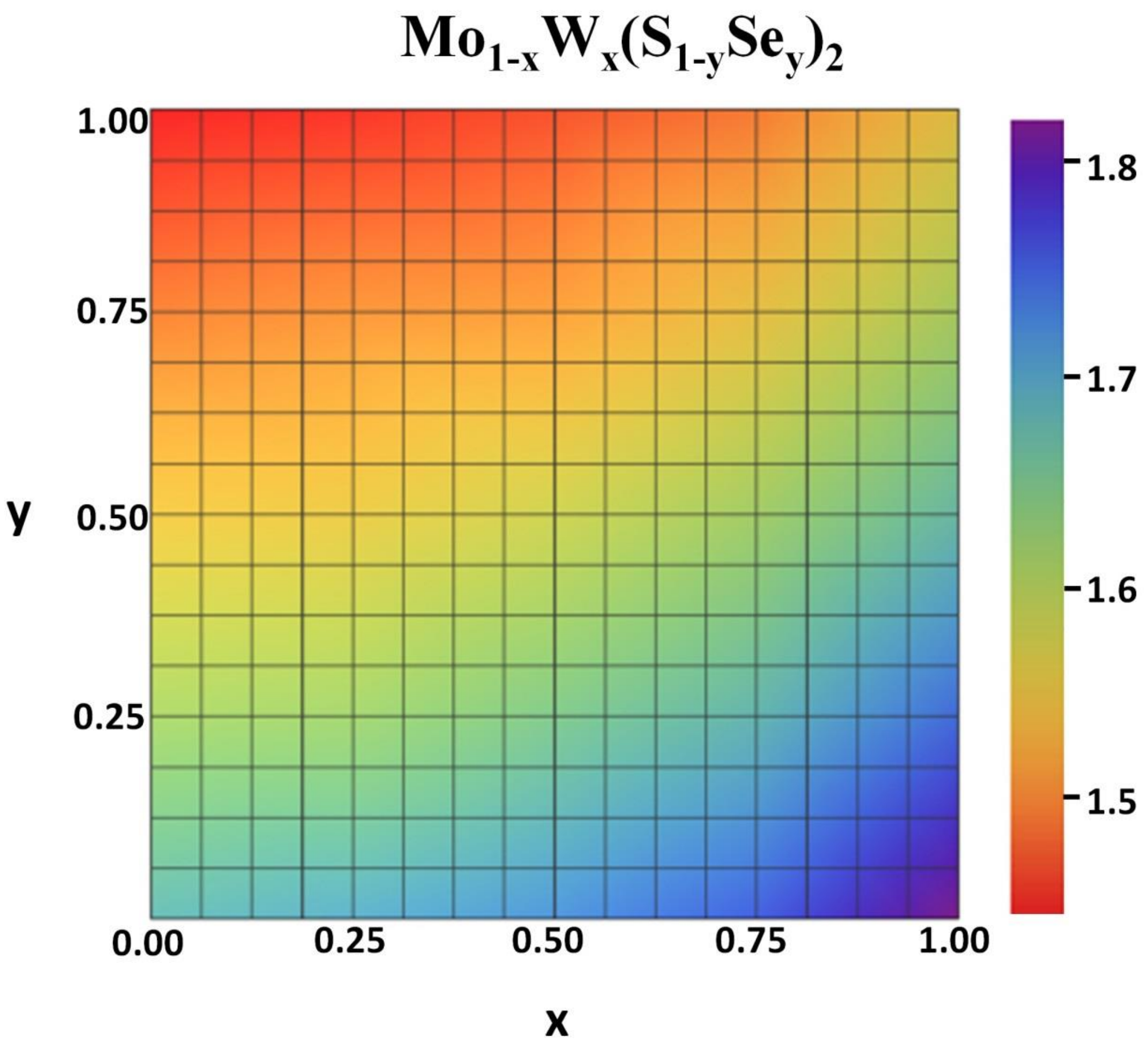

Figure 6. Bandgaps of quaternary TMD alloys $Mo_{1-x}W_x(S_{1-y}Se_y)_2$ as functions of compositions x and y.

Table I. Fitted effective cluster interaction parameters $J_\alpha$ for four TMD ternary alloys.

| $J_\alpha$ (meV) / system | $J_{2,1}$ | $J_{2,2}$ | $J_{2,3}$ | $J_{3,1}$ | $J_{3,2}$ |
|---|---|---|---|---|---|
| $Mo_{1-x}W_xS_2$ | 2.05 | -0.12 | -0.20 | 0.04 | -0.04 |
| $Mo_{1-x}W_xSe_2$ | 1.47 | -0.11 | -0.19 | 0.01 | -0.01 |
| $Mo(S_{1-x}Se_x)_2$ | 2.81 | 1.64 | -0.03 | -0.12 | -0.07 |
| $W(S_{1-x}Se_x)_2$ | 2.73 | 1.61 | -0.21 | 0.04 | -0.26 |

Table II. Cluster correlation functions $\bar{\Pi}_{\alpha}$ in ideally random, SQS (artificially constructed), and locally ordered (CE predicted) TMD ternary alloys at selected compositions.

| $\bar{\Pi}_{\alpha}$ | $\bar{\Pi}_{2,1}$ | $\bar{\Pi}_{2,2}$ | $\bar{\Pi}_{2,3}$ | $\bar{\Pi}_{3,1}$ | $\bar{\Pi}_{3,2}$ | $\bar{\Pi}_{2,1}$ | $\bar{\Pi}_{2,2}$ | $\bar{\Pi}_{2,3}$ | $\bar{\Pi}_{3,1}$ | $\bar{\Pi}_{3,2}$ |
|---|---|---|---|---|---|---|---|---|---|---|
| $Mo_{1-x}W_xS_2$ | | | | | | $Mo(S_{1-x}Se_x)_2$ | | | | |
| x=1/4: | | | | | | | | | | |
| Random | 1/4 | 1/4 | 1/4 | -1/8 | -1/8 | 1/4 | 1/4 | 1/4 | -1/8 | -1/8 |
| SQS | 1/4 | 1/4 | 1/4 | -1/8 | -1/8 | 1/4 | 1/4 | 1/4 | -1/8 | -1/8 |
| Ordered | 0 | 0 | 1 | 1/2 | 1/2 | -- | -- | -- | -- | -- |
| x=1/3: | | | | | | | | | | |
| Random | 1/9 | 1/9 | 1/9 | -1/27 | -1/27 | 1/9 | 1/9 | 1/9 | -1/27 | -1/27 |
| SQS | 1/9 | 1/9 | 1/9 | -1/27 | -1/27 | 1/9 | 1/9 | 1/9 | -1/27 | -1/27 |
| Ordered | -1/3 | 1 | -1/3 | 1 | 1 | -1/3 | -1/3 | 1/3 | 1 | 1 |
| x=1/2: | | | | | | | | | | |
| Random | 0 | 0 | 0 | 0 | 0 | 0 | 0 | 0 | 0 | 0 |
| SQS | 0 | 0 | 0 | 0 | 0 | 0 | 0 | 0 | -1/16 | 0 |
| Ordered | -1/3 | 0 | 5/9 | 0 | 0 | -1 | -1/3 | 1/3 | 0 | 0 |

Table III. Calculated decomposition of formation enthalpies of locally ordered $Mo_{1-x}W_xS_2$ and $Mo(S_{1-x}Se_x)_2$ alloys during the VD, CE, and SR processes.

| | ΔH (meV) | $\Delta H_{VD}$ (meV) | $\Delta H_{CE}$ (meV) | $\Delta H_{SR}$ (meV) |
|---|---|---|---|---|
| | | $Mo_{1-x}W_xS_2$: | | |
| x=0.25 | -5.6 | 0.2 | -5.3 | -0.5 |
| x=0.33 | -7.5 | 0.1 | -7.2 | -0.4 |
| x=0.50 | -8.6 | 0.1 | -7.0 | -1.7 |
| x=0.66 | -7.5 | 0.1 | -7.2 | -0.4 |
| x=0.75 | -5.7 | 0.1 | -5.2 | -0.6 |
| | | $Mo(S_{1-x}Se_x)_2$: | | |
| x=0.33 | -3.4 | 64.4 | -13.8 | -54.0 |
| x=0.50 | -3.5 | 70.2 | -13.9 | -59.8 |
| x=0.66 | -2.9 | 60.1 | -12.8 | -50.2 |